\documentstyle[aps,preprint,12pt,psfig]{revtex}

\begin{document}

\title{On the Influence of Noise on the Critical and Oscillatory Behavior of a Predator-Prey Model: Coherent Stochastic Resonance at the Proper Frequency of the System.}

\author{A.F. Rozenfeld$^{a}$\thanks{Electronic address:
aleroz@inifta.unlp.edu.ar}, C.J. Tessone$^{b,\,c}$\thanks{Electronic address:
tessonec@venus.fisica.unlp.edu.ar}, E. Albano$^a$\thanks{Electronic address:
ealbano@inifta.unlp.edu.ar} and H.S. Wio$^b$\thanks{Electronic address: wio@cab.cnea.gov.ar} \\
$^a$ Instituto de Investigaciones Fisicoqu\'{\i}micas Te\'{o}ricas y Aplicadas (INIFTA),\\ UNLP, CONICET, CIC.
Suc 4, C.C. 16. (1900) La Plata. Argentina.\\
$^b$ Centro At\'omico Bariloche (CNEA) and Instituto Balseiro
(CNEA and UNC), \\
8400-San Carlos de Bariloche, Argentina\\
$^c$ Instituto de F\'{\i}sica de La Plata,\\ C.C. 63, Universidad Nacional de La Plata. (1900) La Plata. Argentina.}
\date{\today}

\maketitle
\newpage
\begin{abstract}
Noise induced changes in the critical and oscillatory behavior of
a Prey-Predator system are studied using power spectrum density
and Spectral Amplification Factor (SAF) analysis. In the absence of
external noise, the population densities exhibit three kinds of
asymptotic behavior, namely: Absorbing State, Fixed Point (FP) and an
Oscillatory Regime with a well defined proper (natural) frequency.
The addition of noise destabilizes the FP phase inducing a transition 
to a new OR. Surprisingly, it is found  that when a periodic signal 
is added to the control parameter, the system responds robustly, without 
relevant changes in its behavior. Nevertheless, the {\em Coherent Stochastic 
Resonance} phenomenon is found only at the proper frequency. Also, a method 
based on SAF allows us to locate very accurately the transition points 
between the different regimes.
\end{abstract}
\newpage

Oscillatory behavior is ubiquitous in all aspects of Nature. For instance, 
time periodic patterns can be found in problems involving the relation 
between different species (competition, predator-prey, coexistence, etc) 
\cite{vol,mur}. Oscillatory behavior can be also found 
in spatial dispersion and competition of living or chemical species 
\cite{mur,oku,chow,fif,nico}, and even in some physiological situations 
(cardiac and circadian rythms) \cite{Win}. 
            
Recently, and related to the interplay between oscillations and noise, 
the phenomenon of stochastic resonance (SR) has been studied in 
different physical, chemical and biological contexts \cite{SR3}. In 
particular, SR has been found to play 
a relevant role in several problems in biology: mammalian sensory systems, 
increment of the tactile capacity, visual perception, effects of low 
fequency and low amplitude electromagnetic fields, etc \cite{biology}. 
Even the problem of coupling among SR units has been studied \cite{adi,wiocas}. 

A related, albeit slightly different phenomenon, is the so called 
{\it stochastic coherent resonance} (SCR) \cite{ref1}. This phenomenon, 
that resembles SR, corresponds to a situation where the system shows 
noise-induced coherent oscillations \cite{ref2} without an externally 
applied signal or a discrete component in its spectrum (in this aspect 
it differs from SR without pariodic forcing as discussed in \cite{ref3}). 
Here we analyze the effect of fluctuations on a lattice gas 
model for a prey-predator system with smart pursuit and escape. Our aim 
is to analyze the possibility that a SCR-like 
phenomenon can occur in such a system.  

Our lattice gas model, which is a variant of the cellular automata 
proposed by Boccara {\it et. al}\cite{boccara}, is defined as follows: 
a lattice site can be either empty or occupied either by a prey  or a 
predator. Double occupancy of lattice sites is forbidden. The system 
evolves according to consecutive cycles: i) competition of species and 
ii) escape-pursuit dynamics.

The competition rules are as follows: {\it (a)}Preys have an offspring 
occupying an empty next neighbor site
with probability $B_{PH}$ (Birth Probability of Preys) in case of absence
of predators within their $V_{RH}$ (visual range of preys).
{\it (b)} Predators can eat a prey that is in their $M_{RP}$ (Movement
Range of Predators) with probability $D_{PH}$ (Death Probability of Preys).
{\it (c)} Predators who have already eaten a prey can produce an offspring
in the site occupied previously by the eaten prey, with probability $B_{PP}$ 
(Birth Probability of Predators).
{\it (d)} Predators can suddenly die with probability $D_{PP}$ (Death
Probability of Predators).

The rules for an escape-pursuit process are as follows: {\it (a)}
Preys calculate the gradient of the density of predators in their
$V_{RH}$ and move into an empty site in the opposite direction.
{\it (b)} Predators calculate the gradient of the density of preys
in their $V_{RP}$ (Visual Range of Predators) and move into an
empty site in that direction. 

We have restricted ourselves to
investigating the dependence of the system on the predator birth
probability ($B_{PP}$), while the remaining parameters are kept
constant, namely: $M_{RP}=V_{RP}=V_{RH}=1$, $D_{PP}=0.25$,
$D_{PH}=1.$ and $B_{PH}=0.5$. The model is studied within a mean
field approach. Further details on the model and results of Monte
Carlo simulations have already been published\cite{alex}.

With the above considerations, the mean field equations for the system can be 
derived evaluating the rates of all processes which may change the species 
densities. Then, it is obtained
\begin{eqnarray}
\partial_t \rho_P &=&\rho_P\cdot \left( B-D\right) \label{pdevol} \\
\partial_t \rho_H &=&\rho_H\cdot \left\{ A-\left(
B+C\right) \cdot \left[ 1-\left( 1-\rho_P\right) ^{(2 M_{RP}+1)^2-1} \right] \right\} \label{hdevol}
\end{eqnarray}
where $\rho _{P}$ and $\rho _{H}$ are the global predator and prey
densities, respectively. Also the rates are given by: {(i)} The
rate of prey reproduction $A=B_{PH}\cdot
(1-D_{PP})^{(2V_{RH}+1)^{2}-1}[1-(\rho _{P}+\rho _{H})^{8}]$ which
comprises the probability of a prey to have an offspring
($B_{PH}$) in a  neighboring empty state (first bracket) provided
the absence of predators in the neighborhood of the progenitor
(second bracket). (ii) The rate of predator reproduction
$B=(1-D_{PP})\cdot B_{PP}\cdot D_{PH}\cdot
[1-(1-\rho_{H})^{(2M_{RP}+1)^{2}-1}]$, which comprises the
probability of the predator to survive $(1-D_{PP})$, the birth
probability of predator ($B_{PP}$), the death probability of a prey
($D_{PH}$) and the probability of existence of a prey in the
visual range of a predator (last term). (iii) The probability of a
predator to catch a prey $C=(1-D_{PP})\cdot (1-B_{PP})\cdot
D_{PH}\cdot [1-(1-\rho_{H})^{(2M_{RP}+1)^{2}-1}]$ is equal to the
previous term $B$, except for the fact that the predator does not
have an offspring and consequently $B_{PP}$ has been replaced by
$(1-B_{PP})$. (iv) Finally, the dying probability of predators is
$D=D_{PP}$.

In order to study the influence of external perturbations on the
system, it is assumed that the control parameter, $B_{PP}$, is
time dependent and has the form
\begin{equation}\label{bppfun}
B_{PP}(t)=B_{PPo} +  Q \, \xi(t) + \varepsilon \cos (\Omega_{\varepsilon } t),
\end{equation}
where $\xi(t)$ is a normalized Gaussian white noise of intensity $Q$ and 
$\varepsilon$ is the amplitude of a periodic external signal with 
frequency $\Omega_{\varepsilon}$. 
In any biological system a parameter such as the Birth Probability 
of Predators (or any other) will not be constant in time. 
Effectively, the environment competition, climate variability, etc,  
will alter the characteristics of the birth rate of all species. Although 
it is true that all the parameters of this system will change in time, as 
a first step in our study, we simplify the analysis by only considering 
modifications on the $B_{PP}$ parameter. 

Clearly, the choice of a Gaussian noise implies the possibility of 
having negative values of $B_{PP}(t)$. Such a case has no physical 
meaning but, as the width of the Gaussian distribution is too small, 
the small negative tail does not affect our results. 

\begin{figure}
\leavevmode
\begin{center}
\hspace{-.5cm}
\psfig{file=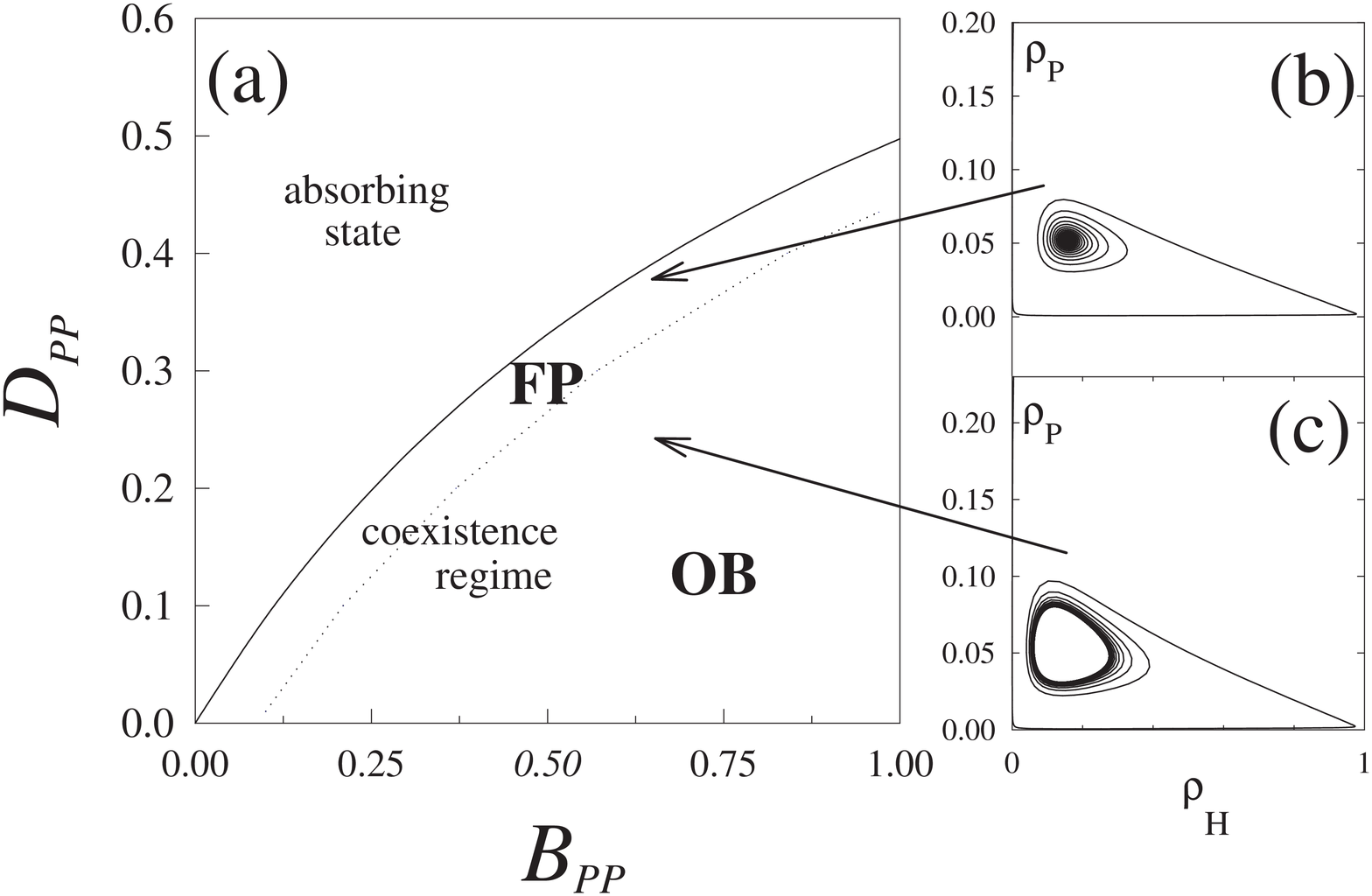,height=12cm,angle=0,silent=true}
\caption{Phase diagram of the prey-predator system as obtained solving 
the mean-field equations in absence of noise. 
(a) Plot of $D_{PP}$ versus $B_{PP}$ showing the 
critical curve (solid line) for the irreversible phase transition
between the prey-predator coexistence regime and the absorbing state
where predators become extinct. The dotted line shows the critical
curve for the transition between the fixed point (FP) regime and the 
oscillatory behavior (OB). Figures (b) and (c) show the dependence
of the population densities for   
the different regimes of the system,
namely the FP and the OB, respectively.}\label{fases}
\end{center}
\end{figure}

Initially, the behavior of the system has been studied in the absence of any
external perturbation, i.e. fixed $B_{PP}$. It was found that the densities 
of species, $\rho_P$ and $\rho_H$ show the following asymptotic behavior
depending on the value of $B_{PP}$: (i) If $B_{PP}<1/3$ the system
evolves toward the extinction of predators, namely $\rho_P=0$, and
$\rho_H=1$. Such a state is an  ``Absorbing Regime'' (AR),
because the system can not escape from it. (ii) If $B_{PP} \ge
1/3$ the final state of the system is a steady regime with
coexistence of prey and predators. However within this regime, 
dynamic behavior changes according the value of $B_{PP}$. Thus, if
$B_{PP}$ is {\em near} the AR phase, $\rho_P$ and $\rho_H$ reach
{\it constant} values, say a ``Fixed Point'' (FP) regime (see Fig.
\ref{fases}). On the other hand, for high enough values of $B_{PP}$, 
the system enters an ``Oscillatory Regime'' (OR), since both
populations exhibit self-sustained oscillations (Fig.
\ref{fases}). It should be noted that the proper frequency of the
system is independent of $B_{PP}$.

Hence, $B_{PP}$ is the control parameter governing the system
dynamics, while $\rho_P$ and $\rho_H$ are order parameters.
Then, it is possible to infer that there are critical values 
$B_{PP}^{1c}$ and $B_{PP}^{2c}$ separating both the AR and the FP, 
and FP and OR, respectively.

The discrete versions of Eqs. (\ref{pdevol}) and (\ref{hdevol}) have 
been solved in order to: (i) determine if the system shows the 
SCR phenomenon; (ii) provide methods for the accurate localization of 
the critical values.

Time series of population densities, $\rho_H(t)$ and 
$\rho_H(t)$, were computed recursively from the discrete mean field 
equations.  Then, the respective temporal self-correlation functions 
$K_H(\tau)$ and $K_P(\tau)$, given by
\begin{eqnarray}
K_H(\tau)=  \langle \rho_H(t) \rho_H(t+\tau) \rangle_t,\hspace{2cm}
K_P(\tau) = \langle \rho_P(t) \rho_P(t+\tau) \rangle_t.
\end{eqnarray}
were calculated. Here $\langle\rangle_t$ represents the time average. 
It is well known that the power spectrum density (psd) is just the Fourier 
Transform of such functions $S_H (\omega ) = {\cal F}\left[ K_H(\tau) \right]$ 
and $S_P (\omega ) = {\cal F}\left[ K_P(\tau) \right]$, respectively. Then, the 
SNR is readily obtained as
\begin{eqnarray}
R_H=10 \log_{10} \left( \frac{S_H^s}{S_H^n}\right) \hspace{2cm}
R_P=10 \log_{10} \left( \frac{S_P^s}{S_P^n}\right),
\end{eqnarray}
where $S_H^s$ ($S_P^s$ ) is the output power at the frequency of
the signal considered, and $S_H^n$ ($S_P^n$) is the output power
of the noisy background at the same frequency, both of them
obtained from the prey (predator) time-series, respectively. From now on, 
we will present  the results for the predator time-series only, as the prey 
ones shows qualitatively the same behavior. Also, the subscripts will be 
omitted to simplify the notation.

It is worthwhile to remark here, that the analysis was made considering
the {\it proper} frequency of the system, $\Omega^* = 0.03$. At this 
frequency $R$ reaches a significant value, in contrast to its value 
at the forcing frequency ($\Omega_\varepsilon$) where $R$ is negligible.
We also verified that the periodic term in Eq. (\ref{bppfun}), does not 
introduce changes in the dynamics of the system. Thus, we will not consider 
in this work any oscillating perturbations on the control parameter.

\begin{figure}
\leavevmode
\begin{center}
\hspace{-.5cm}
\psfig{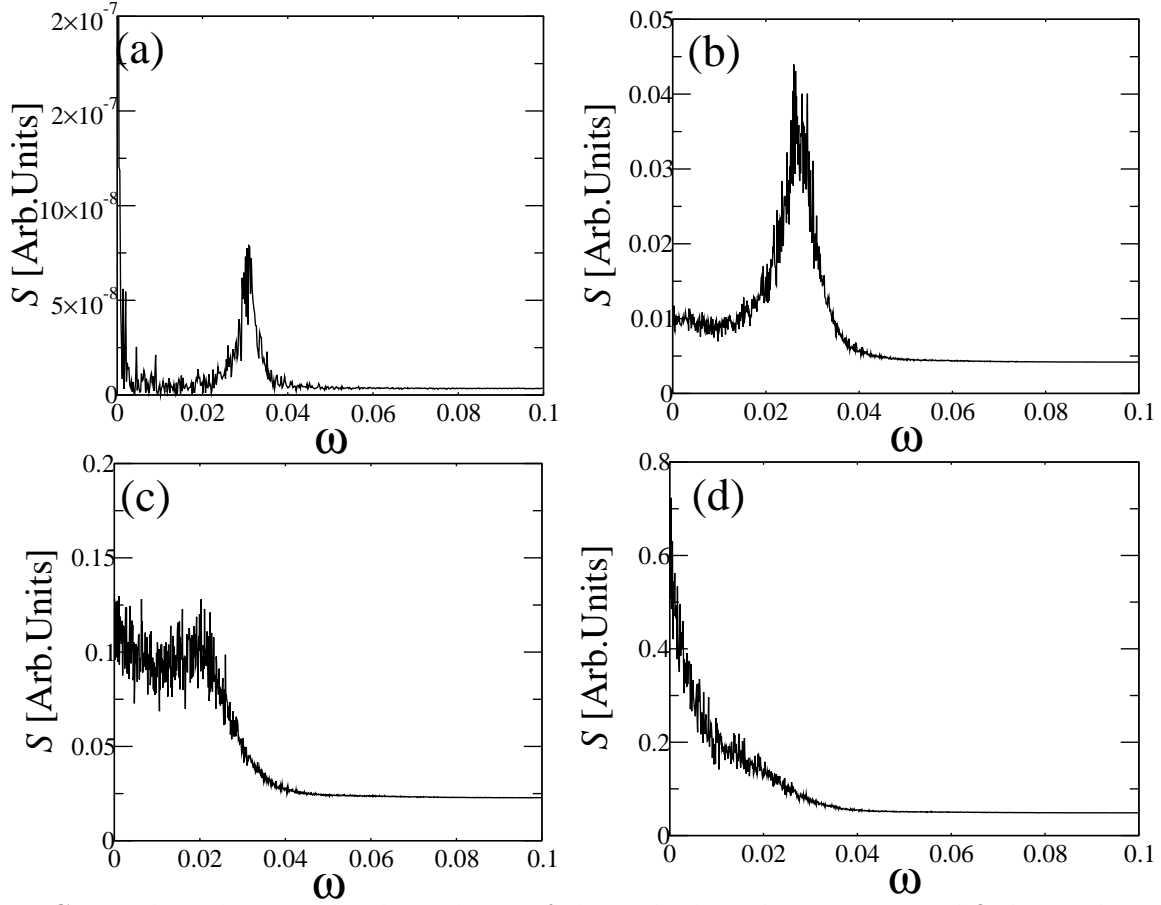}
\caption{Plots showing the dependence of the psd when the noise is modified. 
In absence of noise the system is in the Fixed Point regime since $B_{PPo}=0.42$ 
is kept constant. In each graph, we varied $Q$, the values being: {\bf (a)} 
$Q=10^{-8}$,{\bf (b)} $Q=10^{-2}$, {\bf (c)} $Q=4.45 \cdot 10^{-2}$, {\bf (d)} 
$Q=10^{-1}$.}\label{psd-ptofijo}
\end{center}
\end{figure}

\begin{figure}
\leavevmode
\begin{center}
\hspace{-.5cm}
\psfig{file=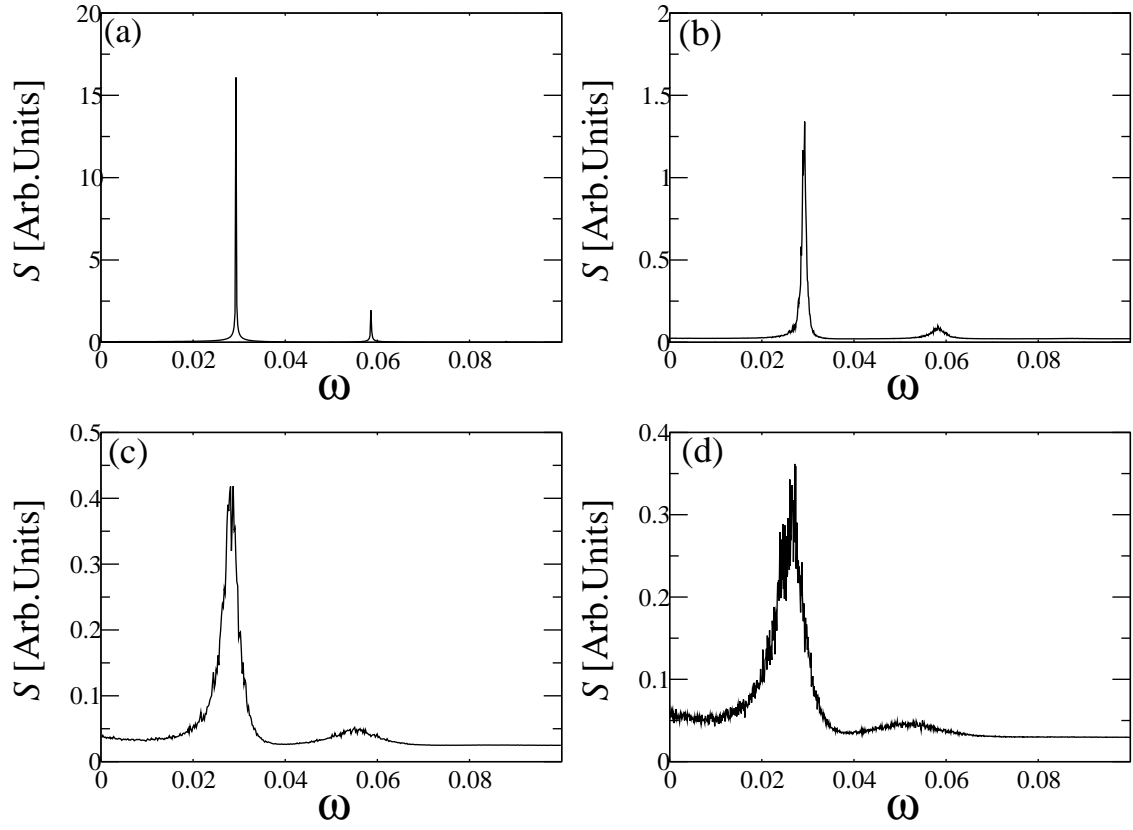,height=12cm,angle=-90,silent=true}
\caption{Plots of $S$ versus $\omega$ corresponding to the time series 
of the predator density. In this case, when $Q=0$ the system is in OR. 
The results are for: {\bf (a)} $Q=0$, {\bf (b)} $Q=10^{-4}$, {\bf (c)} 
$Q=10^{-2}$ and {\bf (d)} $Q=10^{-1}$, keeping $B_{PPo}=0.70$ constant.}
\label{psd-oscil}
\end{center}
\end{figure}

Figures \ref{psd-ptofijo} and \ref{psd-oscil} show the psd when the system 
is in the FP and OR regimes, respectively.
Qualitatively, the results can be understood as follows (see Fig. \ref{psd-ptofijo}): 
if the system is in FP, the psd at the proper frequency is zero when $Q=0$. 
For very low noise strengths ($Q<10^{-4}$), $S(\Omega^*)$ is an increasing 
function of $Q$. However, the situation changes when the noise is strong 
enough ($Q \sim 10^{-3}$). Here, the peak at the proper frequency is hidden 
by the noisy background. Then, it is clear that the function $R(Q)$ 
would have a maximum in the considered range of $Q$.
In contrast, within the OR (see Fig. \ref{psd-oscil}) $S(\Omega^*)$ has 
a finite (non-zero) value in the limit $Q \to 0$. As the noise strength 
grows, such a peak {\it decreases} and broadens. Then, for this case the 
SNR of the system as a function of noise strength is expected to be a 
monotonic decreasing function of such a parameter.

\begin{figure}
\leavevmode
\begin{center}
\hspace{-.5cm}
\psfig{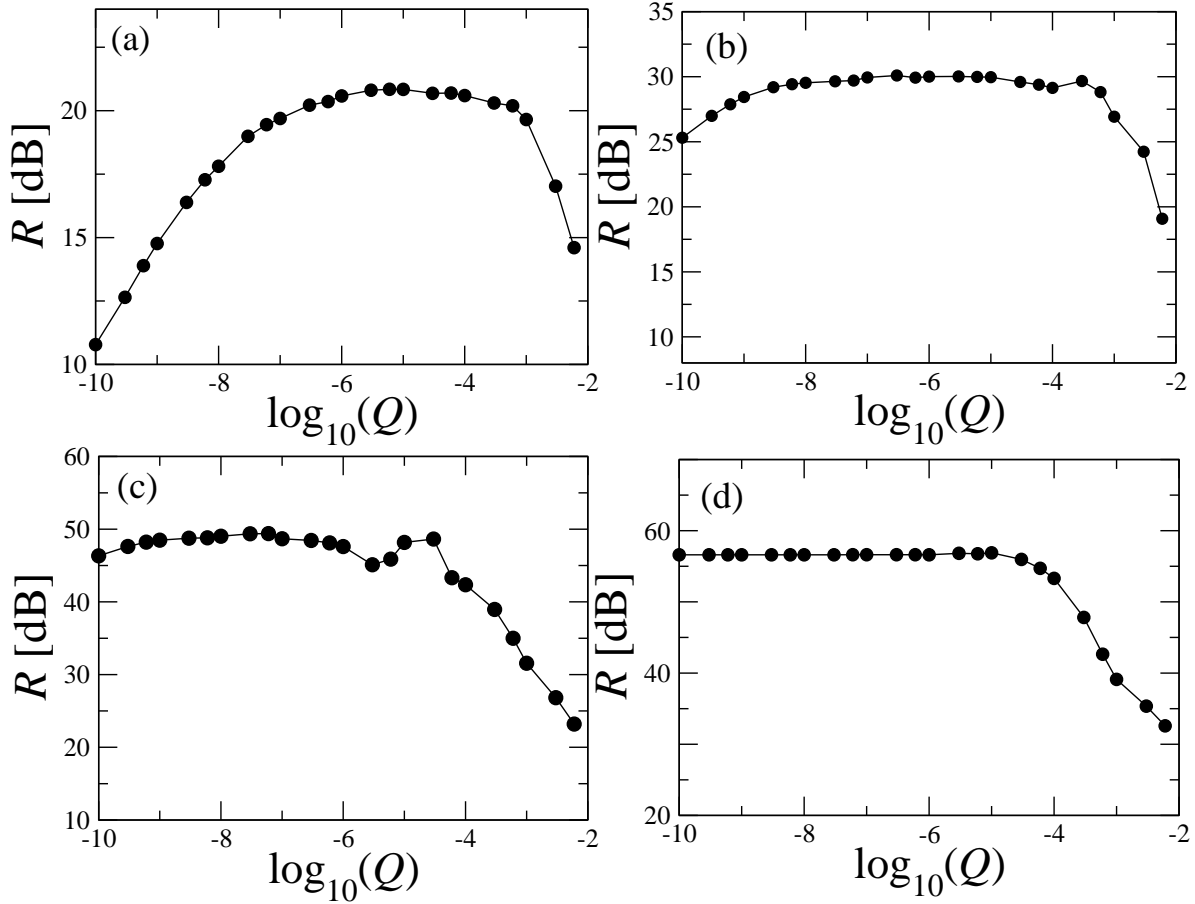}
\caption{Plots of the Output SNR as a function of noise strength, $Q$, obtained 
for different values of $B_{PPo}$. A SCR phenomenon is apparent when $B_{PPo}$ 
is {\em within} FP regime, but it disappears if the system is within OR.  
{\bf (a)} $B_{PPo}=0.42$, {\bf (b)} $Q=B_{PPo}=0.45$, {\bf (f)} $B_{PPo}=0.47$, 
{\bf (d)} $B_{PPo}=0.50$.}\label{snr-Q}
\end{center}
\end{figure}

In the following, a method to locate the phase transition between FP 
and OR regimes, by means of the calculation of the SNR, is developed. 
Figure \ref{snr-Q} shows plots of the SNR as a function of $Q$ for different 
values of $B_{PPo}$. The SCR phenomenon is only apparent for values of 
$B_{PPo} \le 0.47$. Such a dependence on $Q$ is the main characteristic 
of the system for the FP regime. When $B_{PPo}>0.47$ the SNR is a decreasing 
function of $Q$, which is the expected behavior when the system is in OR. The 
value of the parameter $B_{PP}^{2c}$ where the SCR phenomenon disappears can 
readily be identified with the critical point. 
Hence, our best estimate for such a transition point is given by 
$B_{PP}^{2c} \cong 0.47 \pm 0.01$.

\begin{figure}
\leavevmode
\begin{center}
\hspace{-.5cm}
\psfig{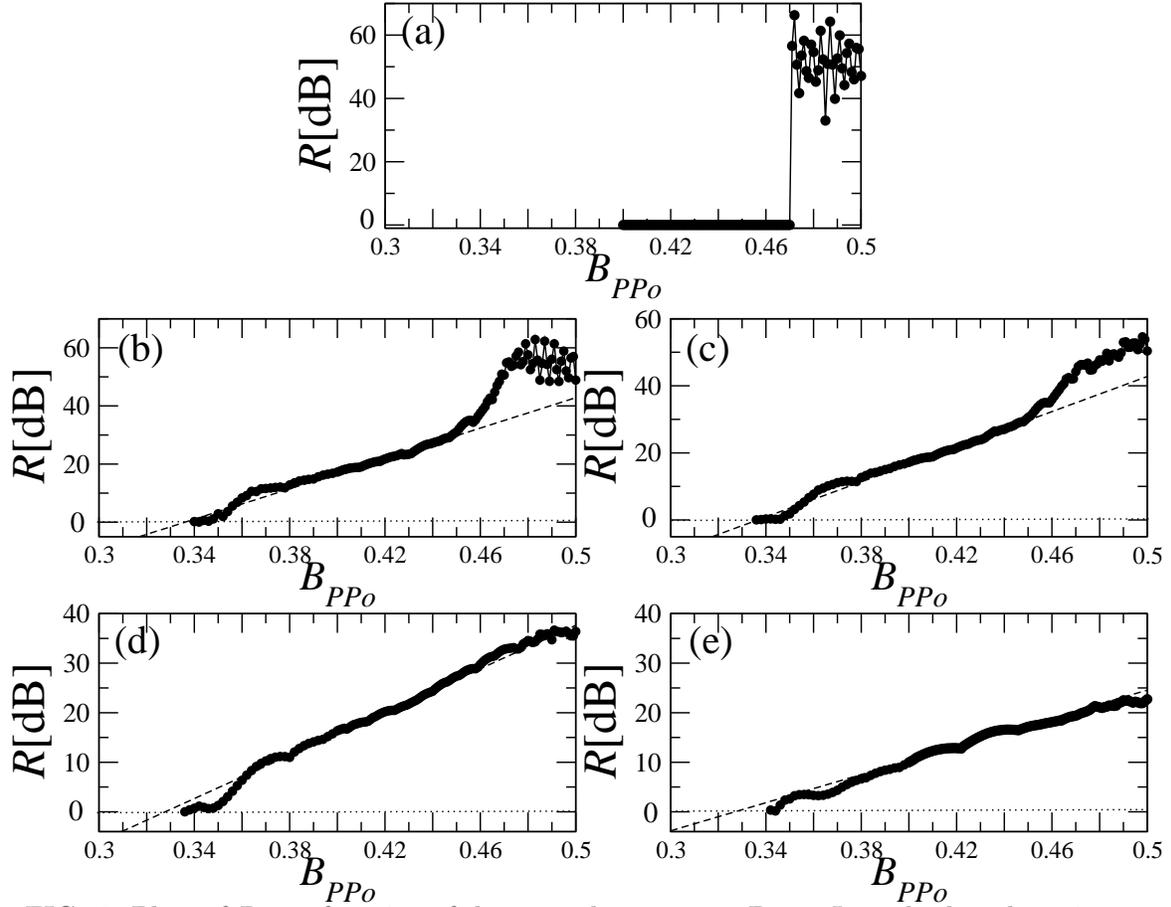}
\caption{Plots of $R$ as a function of the control parameter $B_{PPo}$. In each 
plot, the noise strength is kept constant, as follows: {\bf (a)} $Q=0$, {\bf (b)} 
$Q=10^{-5}$, {\bf (c)} $Q=10^{-4}$, {\bf (d)} $Q=10^{-3}$ and {\bf (e)} $Q=10^{-2}$.}
\label{snr-BPP}
\end{center}
\end{figure}

Figure \ref{snr-BPP} shows the dependence of SNR as a function of
$B_{PPo}$, for some fixed values of $Q$. In the absence of
noise the critical point can easily be obtained from the plot
(Fig. \ref{snr-BPP}(a)), yielding $B_{PP}^c \cong 0.47\pm 0.01$, in
agreement with the value obtained through the method discussed
above.  

A surprising consequence of the addition of noise is that, as can be 
seen in Figs. \ref{snr-BPP}(b)-(e), even for control parameter values 
$B_{PPo}<0.47$ the system oscillates at the proper frequency. This indicates 
that the FP phase is unstable, and even a small perturbation drives it into 
an oscillatory behavior.
Also, the plots of $R$ vs $B_{PPo}$ exhibit linear
behavior for lower values of the parameter, e.g $B_{PPo}<0.46$.
Extrapolations by linear regression to the limit
$R\to 0$ give $B_{PP}^{1c}a \cong 0.330 \pm 0.005$ which is the
critical point for the irreversible transition from the
coexistence regime to the AS. This figure is in excellent
agreement with our previous estimate $B_{PP}^{1c}=1/3$ obtained
simply solving the mean field equations in the absence of noise, as
discussed earlier. 

Subsequently, for small noise values $Q\leq 10^{-4}$ (Figs. \ref{snr-BPP}(b)-(c)), 
the onset of a shoulder is clearly observed, say for
$B_{PPo} > 0.47$. As will be discussed later, this can be associated to another 
phase transition between intrinsically different oscillatory regimes. 

In the search of a method to characterize these two different oscillating 
phases we make use of the fact that, although this is an extremely non-linear 
system, the time series $\rho_H$ and $\rho_P$ are composed only by a single
frequency (the proper one) and its harmonics. Supported by this
numerical evidence, it is assumed that for long times, the
solution of Eqs. (\ref{pdevol}) and (\ref{hdevol}) approaches
a periodic function, which will be of Floquet-type. Then, the
corresponding stationary self-correlation function
($K_{as}(\tau)$), can be decomposed in a Fourier series, i.e.
\begin{equation}\label{hanggieq}
K(\tau) = \sum_{n=1}^{\infty} A_n \sin (n \Omega \tau) + \sum_{n=0}^{\infty} B_n \cos (n \Omega \tau).
\end{equation}
It is expected that in the FP regime, all the coefficients (with
the exception of $B_0$) will vanish. Also, in OR, at least the
terms $A_1$ and $B_1$ (related with the amplitude of oscillations) must 
be non zero. Hence, measuring the value of those coefficients as a 
function of $Q$ and $B_{PP}$ it may be possible to find the phase transition. 

Once again, this is a SCR phenomenon, as analyzed within the 
framework used in Ref. \cite{saf-hanggi}, where the characterization of 
SR is done by means of the Spectral Amplification Factor (SAF). Following this 
course, we define ${\hbox{\tt W}}$, the {\it degree of oscillation} 
(this will indicate how much inside the OR is the
system), as the inner product between $K_{as}(\tau)$ and the first harmonic 
of Eq. (\ref{hanggieq}), which gives
\begin{equation}
{\hbox{\tt W}}=A_1^2+B_1^2.
\end{equation}

\begin{figure}
\leavevmode
\begin{center}
\hspace{3cm}
\psfig{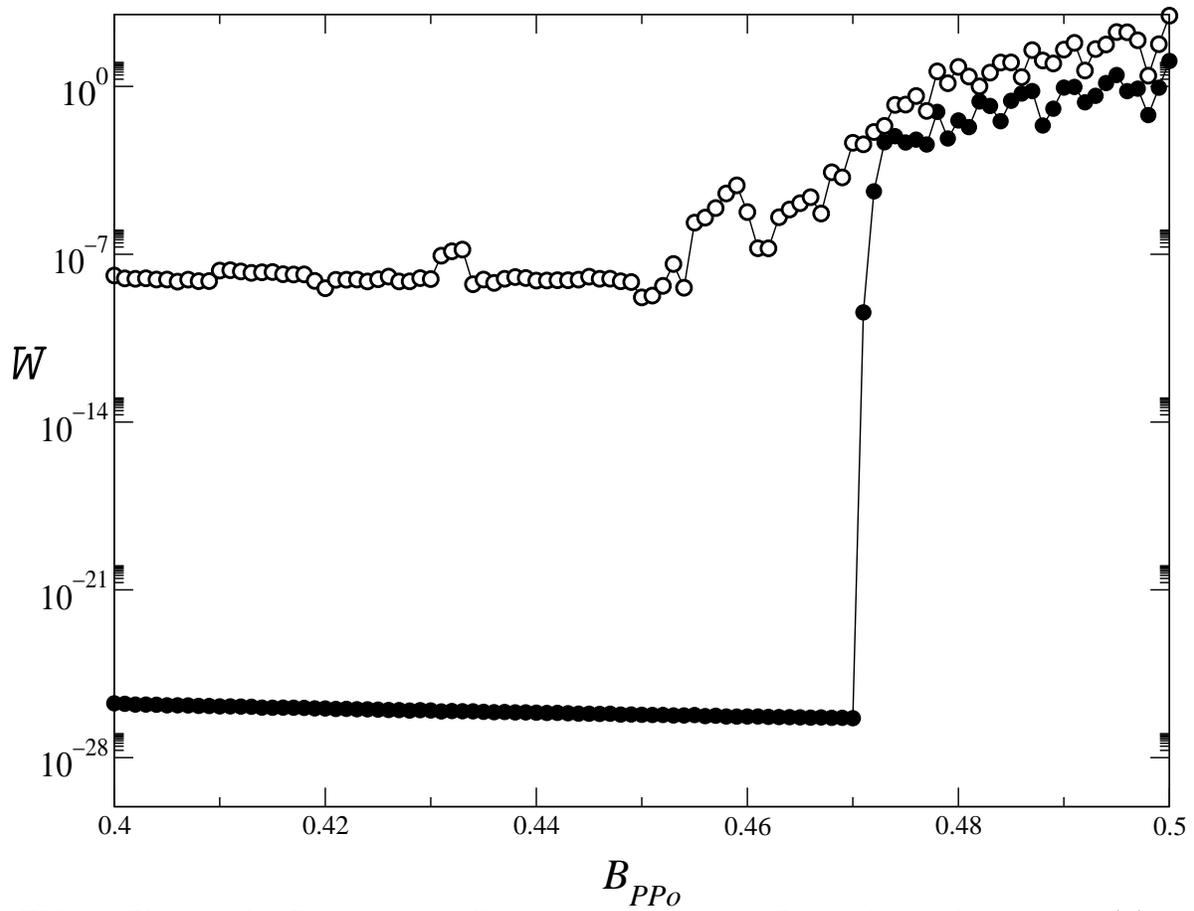}
\caption{Plots of the {\it Degree of Oscillation} as a function of $B_{PPo}$, 
obtained for $Q=0$ {\bf (a)}, and $Q=10^{-5}$  {\bf (b)}. In both cases, the 
phase transition between FP and OR can be observed. Notice that such a 
transition {\it is shifted} in the presence of a small noise strength. }
\label{saf-BPP}
\end{center}
\end{figure}

Figure \ref{saf-BPP} shows plots of ${\hbox{\tt W}}$ 
as a function of the control parameter $B_{PPo}$, in the absence (filled circles) 
and presence (empty circles) of noise. It is easy to distinguish the following: 
(i) In the absence of noise the FP regime is clearly observed in the Figure, and a 
value $B_{PP}^{2c}=0.47$ is found, in 
excellent agreement with the results previously discussed in connection 
with the behavior of SNR vs. $B_{PPo}$ (Fig. \ref{snr-BPP}). (ii) In the presence of 
noise it can be seen that the value of {\tt W} increases for $B_{PPo} < 0.45$. 
Then, a new oscillatory regime is apparent where {\tt W} is independent of 
$B_{PPo}$, and such a behavior arises instead of the FP phase, because of the 
applied low-intensity noise. This phenomenon can be understood in terms of a 
noise induced phase transition. On the other hand, the usual OR where {\tt W} 
monotonically increases with $B_{PPo}$, is obtained for $B_{PPo} \geq 0.45$. 
In fact, the value of $B_{PPo}$ for which the system enters into the usual 
OR, suffers a shift due to the presence of noise \cite{nit}.  Effectively, it 
changes from $B_{PPo}(Q=0)=0.47$ to $B_{PPo}(Q=10^{-5})=0.45$. 

The former study of {\tt W} behavior allows us to give further
support to our discussion of the results shown in Fig. \ref{snr-BPP}. 
Substracting the linear behavior, the
following estimates for the phase transition are obtained: 
$B_{PPo} (Q=10^{-5}) \cong 0.461 \pm 0.005$, and 
$B_{PPo} (Q=10^{-4}) \cong 0.453 \pm 0.005$. Finally, when $Q$ 
increases (Figs. \ref{snr-BPP}(d) and \ref{snr-BPP}(e)), the linear 
dependence on $B_{PPo}$ holds for a wider range in such a variable and 
consequently, the phase transition is more difficult to detect.
This is because the system is strongly oscillatory even if (in absence of 
noise) $B_{PPo}$ is within the FP regime. In fact,  the linear tendency in 
the function $R(B_{PPo})$ observed for small values of $Q$ is broken when 
the parameter is set within the OR. Also, when the OR is masked by the 
response to a larger noise, the linear dependence is not broken but the 
critical point can not be located anymore. Furthermore, from 
Figs. \ref{snr-BPP}(a) and \ref{saf-BPP} one observes that in the absence of 
noise the transition $FP \leftrightarrow OR$ is abrupt resembling first 
order behavior. However, when the noise is added the transition between the 
two oscillatory regimes not only becomes shifted, but also rounded 
(Figs. \ref{snr-BPP}(b) and \ref{saf-BPP}).

Summarizing, the effect caused by external noise on a
Prey-Predator system, with smart pursuit evasion is studied by
means of {\em SNR} and {\em Degree of Oscillation} techniques. In
order to perform the study, the behavior of the system is analyzed when 
a small amount of white noise is applied over a control parameter
($B_{PPo}$). It is found that if (in the absence of noise) such 
a parameter is
in FP, a Coherent Stochastic Resonance phenomenon arises. In fact, only
the {\it proper} frequency becomes strongly enhanced, while any 
externally applied oscillatory signal is not amplified. Nevertheless, 
such an enhancement arises only for a (intermediate) range of noise
strength ($Q \sim 10^{-6}$) and it vanishes in two distinct
limiting cases: (i) when noise is not applied and (ii) when the noise
is too large ($Q > 10^{-3}$). It should be remarked that this
resonant phenomenon disappears if the system is driven to the OR,
tuning the control parameter. Then, it is possible to locate,
looking at the dependence of SNR as a function of noise, the
critical point at which the phase transition between the FP and
the OR occurs.

In the presence of noise, the FP phase is replaced by a new oscillatory regime 
where the amplitude of oscillations remains constant, allowing us to assert 
that the FP phase is less stable than the OR one. The dependence of SNR as a 
function of $B_{PPo}$ 
shows linear behavior and the extrapolation of such a line intersects 
the SNR axis at $B_{PPo}=0.330 \pm 0.005$. This value is the critical 
point where the irreversible phase transition to the AS occurs. Thus, 
this method allows us to find {\bf both} phase transitions: (i) the 
irreversible one between th Coexistence Regime and the AS, and in the absence 
of noise (ii) the reversible one between the FP and the OR regimes, respectively.

We have also defined the {\it Degree of Oscillation} as the corresponding 
terms of the first harmonic in the Floquet-type asymptotic expansion of the
self-correlation function of the population time series.
This characterization, provides us with an alternative method to locate the 
phase transition between FP and OR.

Furthermore, the abupt (first-order like) transition between the FP and the 
OR in absence of noise becomes clearly rounded (second-order like) when 
noise is applied. These results point out that the effect of 
noise is not merely restricted to the shift of the critical point,
but instead that the whole nature of the transition changes.   

We expect that the findings reported in this work will contribute to the 
understanding of resonant effects and critical behavior in actual 
competitive population systems, since in nature they are exposed to 
different sources of external noise. \\ \\

{\bf Acknowledgments}

The authors want to thank V. Gr\"unfeld for a critical 
reading of the manuscript.

%*********************************************************
\newpage

%Lista de Figuras

\end{document}